\documentclass{article}
\usepackage[utf8]{inputenc}
\usepackage{authblk}
\usepackage{hyperref}
\usepackage{xcolor}

\newcommand\blfootnote[1]{%
  \begingroup
  \renewcommand\thefootnote{}\footnote{#1}%
  \addtocounter{footnote}{-1}%
  \endgroup
}

\title{Crowd Control in Plazas Constrained to Social Distancing}
\author[1,2]{{\'A}lvaro Gonz{\'a}lez Garc{\'i}a$^{*}$} 
\author[1]{James L. Mart{\'i}n R.}
\author[3]{Alessio Caciagli}
\affil[1]{Van ’t Hoff Laboratory for Physical \& Colloid Chemistry, Department of Chemistry \& Debye Institute, Utrecht University, The Netherlands}
\affil[2]{Laboratory of Physical Chemistry, Department of Chemical Engineering \& Chemistry \& Institute for Complex Molecular Systems (ICMS), Eindhoven University of Technology, The Netherlands}
\affil[3]{Cavendish Laboratory, 19 JJ Thomson Avenue, Cambridge CB3 0HE, UK}

\date{\today}

\usepackage[numbers]{natbib}
\usepackage{graphicx}

\begin{document}
\maketitle
\blfootnote{${^*}$ to whom correspondence should be addressed: a.gonzalez.garcia@tue.nl}
\section*{Abstract}
We present a simple and versatile method for calculating the maximum capacity of public spaces constrained to social distancing, following the recommended measures of the WHO due to the COVID-19 pandemic outbreak. This method assumes a minimum required distance of two meters between persons and is tested in four actual plazas. Additionally, we estimate public space capacity in case of public events, such as markets or concerts. We believe our method will directly provide guidelines for crowd control in the so-called `new normality’, serving as a first step for future and more accurate developments. 
\vspace{5pt}
\hrule
\vspace{6pt}
\clearpage
The relaxation of the COVID-19 pandemic lock-down measures across many countries \cite{noauthor_australias_2020,correspondent_coronavirus_2020,welle_wwwdwcom_coronavirus_nodate} prompts for the swift search of simple methods that provide crowd control guidelines accounting for the recommended social distance established by the World Health Organization (WHO) \cite{who_2020}. There is a strong need to recover social interaction as efficiently and safely as possible, within the reasonable limits imposed by current social distancing measures. So far, public demonstrations in plazas have revealed both the best and the worst of social behavior during this crisis \cite{YoutubeVideo}. Moreover, even after this somehow predicted COVID crisis, re-emergence of other corona-related infections should be anticipated \cite{cheng_severe_2007}. This prompts the need of a strategy to estimate the number of people that can fit in a public space in compliance with social distancing rules, in order to promote the re-emergence of public events conducted in a safe manner. This strategy would not only be of immediate relevance but also of use in the long-run.

Here, we present a simple yet versatile method that provides an estimate of the capacity of public spaces (with a focus towards town plazas) compatible with social distancing. We model the occupancy of a public space as a very loose random packing (VLRP) of hard discs in confinement. Although the method is applicable to any public -and potentially private- space, we focus on plazas. We stress that we do not aim to calculate the optimal packing of people. On the contrary, we choose a purely stochastic approach in order to mimic human behaviour and to allow for a certain degree of personal movement, vital during public events. We also account for the presence of confinement barriers as measures to control crowd occupancy, treating them as a \textit{depletion zone} \cite{lekkerkerker_colloids_2011} around the public space's perimeter. This depletion zone naturally arises due to the excluded area imposed by the outer barrier (and potentially any inner obstacle) to the hard discs. That is, we remove the possibility of people being placed in the vicinity of the perimeter of the square. The purpose of this simple, modular and accessible method is to offer everyone a way to take back our public spaces while respecting the safety guidelines. We envision that this will eventually accelerate the process of return to social normality.

We consider a public space of interest (public square) as a closed polygon. Its geographical position -latitude and longitude- is freely accessible using Google Maps. The polygon is then constructed by drawing, via point-clicking \cite{noauthor_plotting_nodate}, a set of enclosed lines on the map. Despite the limitations of this method (such as the necessity of manual annotation and drawing imprecision), we believe that this approach is simple enough to be accessible not only by scientists, but by any potential user (as explained at the end of this Communication).
We model each person as a two-dimensional hard disc with radius $R=2.0\,\mathrm{m}$. This value is slightly larger than the minimum distance suggested by WHO \cite{who_2020}. The reason of this choice is two-fold. Firstly, our calculations provide a lower estimate as obstacles in the square are not accounted for. Secondly, we believe that a social distance of $1.0\,\mathrm{m}$ \cite{who_2020} will most likely result in social discomfort, particularly in the period directly following the relaxation of lock-down measures.
The crowding of the public space is simulated by random sequential addition (RSA) \cite{zhang_precise_2013,torquato_random_2006, chen_random_2017,widom_random_1966} of 2D discs into the polygon. From a purely physical perspective, we have developed a method to generate VLRPs of discs constrained to an arbitrary container.
All computations are conducted using Wolfram Mathematica \cite{mathematica_2020}[V. 12]. The script is available as Supplemental Information (SI). 

\begin{figure}
  \centering
  \includegraphics[width=0.5\textwidth]{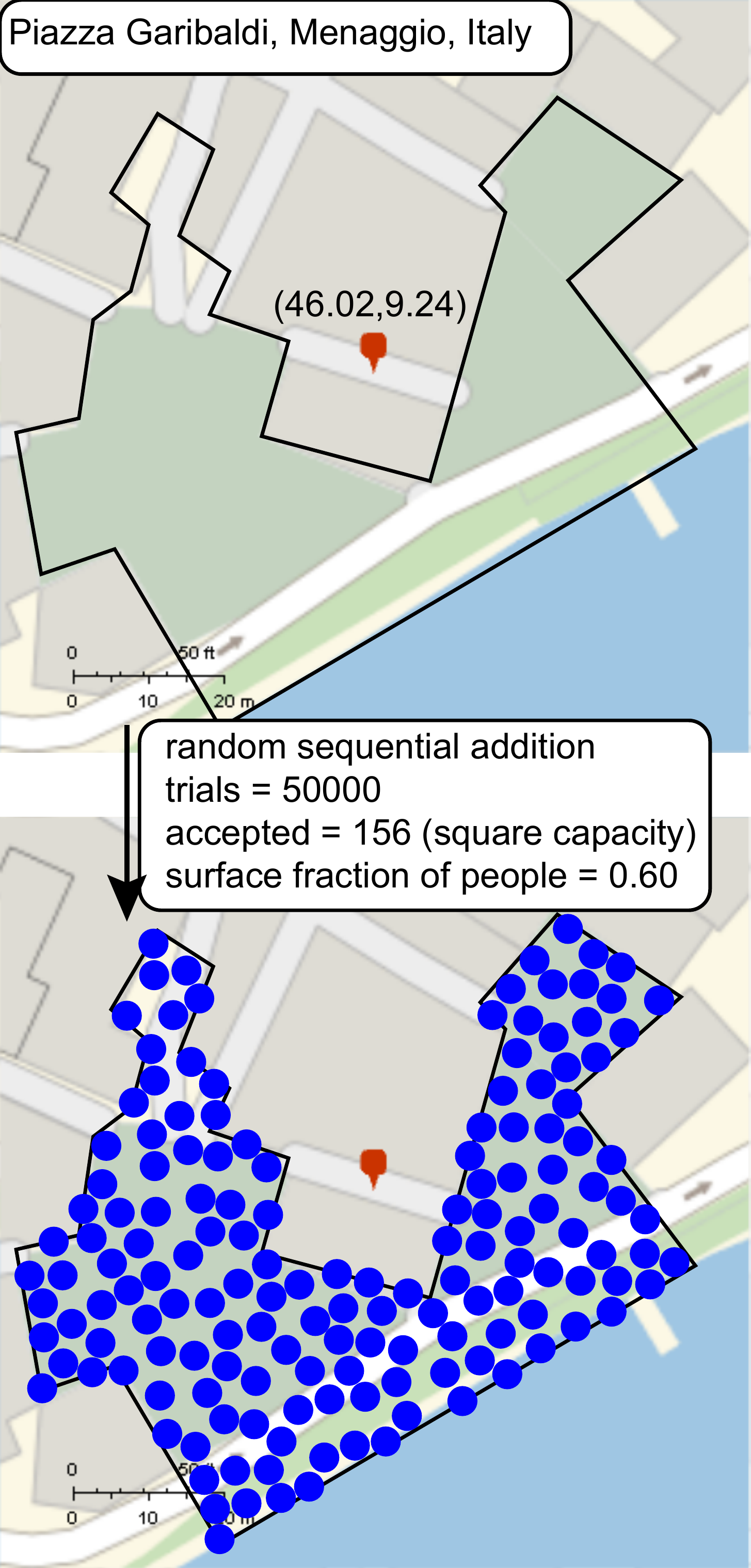}
  \caption{Estimation of the capacity of a public space (Piazza Garibaldi, Menaggio, Italy) using RSA. The top panel shows the geo-localization of the public space in Google Maps (note the geo-coordinates, required as an input). A polygonal line drawn by point-picking on the map defines the confined space of the Piazza: the simulation box for the RSA. The bottom panel shows the Piazza at maximum capacity as obtained by RSA. Each person is represented by a 2D blue disc. We performed $ 5 \cdot 10^{4}$ person insertion trials, resulting in 156 accepted positions (corresponding to an area fraction of $\eta=0.6$). The number of accepted discs represents the capacity of the public space in compliance with social distancing measures. }
  \label{fig:showcase}
\end{figure}

Our method is illustrated in Fig. \ref{fig:showcase}. We intentionally pick Piazza Garibaldi due to its non-trivial shape, which showcases the versatility of our approach. The bottom panel shows a VLRP obtained by RSA after $ 5 \cdot 10^{4}$ attempted disc insertions. The Piazza displays an almost optimal usage in compliance with social distancing measures, as exemplified by the presence of discs even in the top-left small pocket, a region not readily accessible. We estimate a capacity of 156 people for this Piazza, corresponding to an area fraction of $\eta = 0.60$. Remarkably, and perhaps due to the non-trivial plaza shape, $\eta$ is greater than the expected value upon RSA of discs in 2D ($\eta\approx 0.54$)\cite{torquato_random_2006}. This is an intriguing topological feature yet to be explored. Moreover, note that we allow overlap of the discs with the container walls: a person inside the square is not required to have all of its spread area within the square.

We note that the polygonal space is assumed not to have any exclusion zones in its inner region. As a consequence, our method is currently not suited for open spaces with plenty of obstacles. The addition of exclusion zones inside the polygon can, however, be easily achieved by calculating the Connolly surface (also known as Lee-Richards molecular surface) related to the obstacle of interest. Such a surface approximation is quickly implemented and results in a union of discs. Disc-obstacle overlaps can then be efficiently evaluated, keeping computational times within reasonable limits even when multiple objects are present\cite{totrov_contour-buildup_1996,ertl_fast_2000,vorobjev_sims_1997,moon_fast_1989}. This is planned in a future implementation. Once again, as this rod-shape results from the union of two discs, extending our method to account for binary particle mixtures (singles = discs, couples = rods) is straightforward.

We do not limit ourselves only to the calculation of the capacity of public spaces. We also consider how this capacity is reduced due to public events such as markets, concerts or fairs. We assume the presence of a confinement barrier as a way to separate the public space from its surroundings. Such a barrier could be made, for instance, of railings. This could be a possible practical solution towards controlling the number of people allowed in the space. \\
The surface area available during a social event is calculated based upon the concept of excluded volume between particles \cite{gray_theory_2011}. The presence of a barrier excludes a given area close to the perimeter of the public square. In other words, the presence of the barrier leads to a depletion zone\cite{garcia_polymer-mediated_2019} for people around the perimeter of the square. We approximate this depletion zone by calculating the Connolly surface of the boundary. This results in a series of hard discs centered at subsequent points on the perimeter of the square. Each disc leads to an excluded area for the particles to be inserted. This allows us to approximate depletion zones of arbitrary 2D-shapes, provided the hard disc construction is sufficiently fine grained.
We employ RSA as before to obtain the reduced capacity of the public space of interest. As practical case studies, we focus on two plazas: an almost rectangular, small square in the town of Los Silos (Canary Islands, Spain), and the large, famous, and half-moon shaped Puerta del Sol in Madrid, Spain. Fig. 2 shows the estimation of the plazas' capacity in the presence and absence of social events. To give a rough estimation of the capacity of the plaza capacity accounting for obstacles, we superimpose the latter obtained by inspecting the satellite view of the map. We then remove discs whose center of mass overlaps with any of the obstacle. Albeit this is certainly a crude approximation, we plan to add full-fledged interior obstacles in a future implementation, as already mentioned. Moreover, we could also account for the presence of couples who are allowed to walk in proximity to each other. They could, for instance, be defined as a rod with a short aspect ratio (as shown in Fig.2, column 3). 

In case of small towns like Los Silos, open markets used to be a common occurrence, and we hope they will be again. However, the novel social distancing must be respected. In this case, the vending stands will act as a natural separation barrier between the square and its surroundings. Vending stands in Tenerife normally have a size of 3x3m$^2$, and are placed with a gap of approximately $1\,\mathrm{m}$ between stands \cite{dorta_macarena_nodate}. In order to achieve a real-life realisation of crowd control, we propose that all market stands are positioned at the boundary of the square and oriented towards its geometrical centre. In this way, local authorities will be able to easily control the number of people in the square, and the vending stands will serve as a barrier themselves, isolating the square and the event from its surroundings. Note, however, that in our script we can set any barrier size therefore mimicking, for instance, (transparent) shields that enable people on each side of these screens to come closer than $2$m.

The stands are placed following the generation of a mesh of points distributed on the boundary. Uniformity is achieved by distributing the mesh points at equal distances across the arc length of the boundary. The optimal number of stands is obtained as the maximum number of mesh points that satisfy the condition of minimum distance (here taken as the $1\,\mathrm{m}$ gap between stands). After accounting for each stand's area, we remove stands that present an overlap. This results in a rough estimate of the maximum number of stands that can be placed within the event, while accounting for their physical area and the required minimum distance between stands. We define the square efficiency as the number of consumers per stand. By examining its value for the plaza in Los Silos, we observe that maximizing the number of stands minimizes the number of customers per stand: a safer scenario than having fewer stands packed with many customers. 
\begin{figure}
  \centering
  \includegraphics[width=0.9\textwidth]{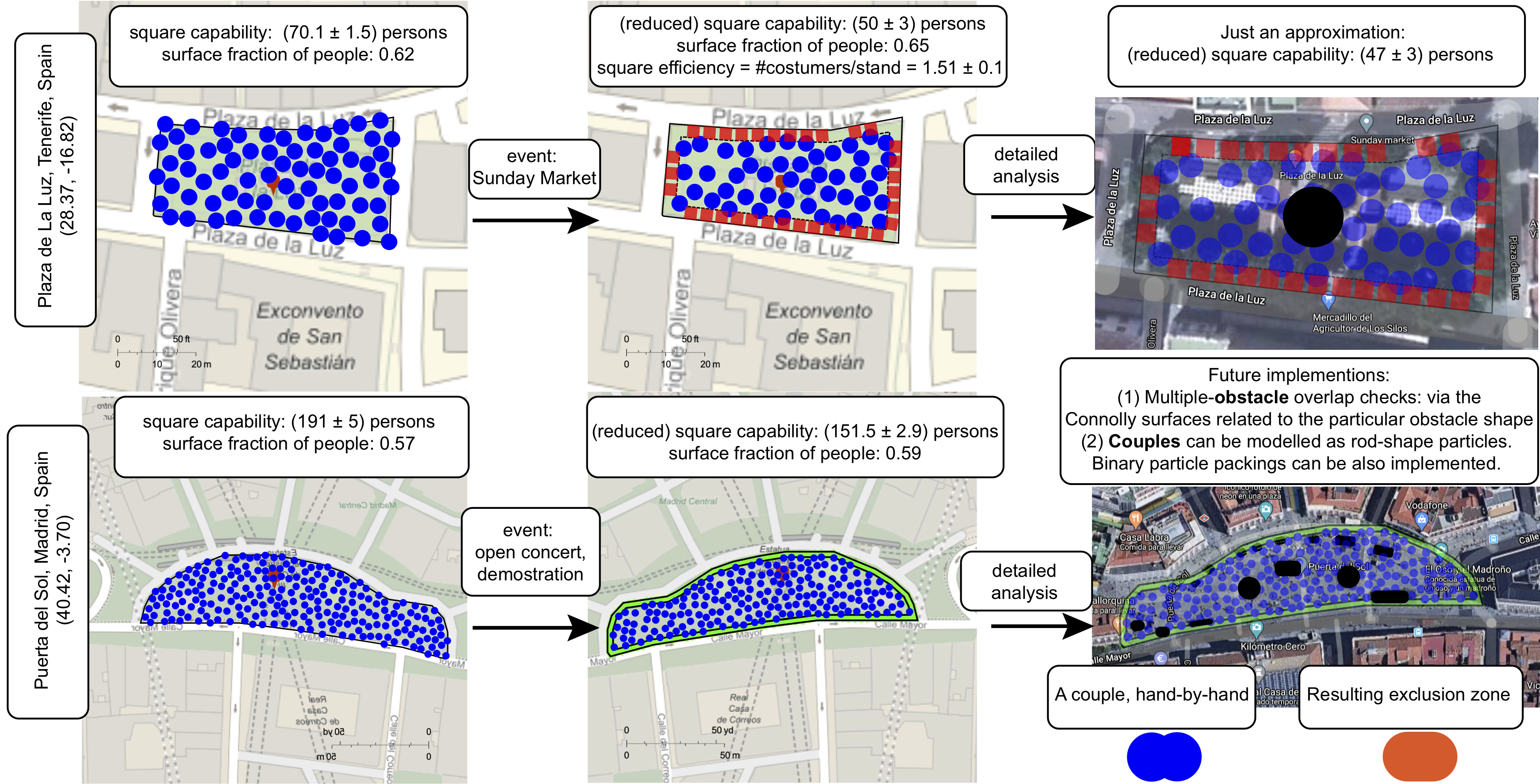}
  \caption{Estimation of the capacity of two plazas (Plaza de la Luz, Los Silos, Canary Islands, and Puerta del Sol, Madrid, Spain) in presence and absence of hypothetical social events. The results for the two plazas are shown in the two different rows.
  First column: estimated square capacity without social events (no barriers); each blue disc corresponds to a person with its corresponding safety radius. Second column: estimated square capacity with social events. For Puerta del Sol, the excluded area due to the event's barrier (depletion zone) is depicted in green. For Plaza de La Luz, we show instead the vending stands acting as a barrier, depicted in red.
  Third column: estimated square capacity accounting for obstacles. In Plaza de La Luz, the correction is obtained by simply superimposing the disc defined by the marquee (black), and checking whether the center of masses of the individuals overlaps with the obstacle.}
  \label{fig:fig2}
\end{figure}

Our simple and versatile approach results in: (a) a rational estimate of the capacity of public spaces subject to social distancing (b) an estimate of this capacity in case of a social and commercial event (case study: Sunday Markets) (c) a quantification of the most efficient plazas in terms of the number of people per stand. This approach will be tested for the Mercado Artesanal del D{\'i}a de Canarias (a regional craftsmen's market) in the municipality of Los Silos, Tenerife (Spain), in close collaboration with two local craftswomen \cite{noauthor_maite_nodate,noauthor_pisando_nodate}.

On top of its practical achievements, our method also well describes the physics behind very loose random packings in two dimensions. This is illustrated in Fig.3, where we apply our method to Kuybyshev Square, the largest square in Europe. For this humongous plaza, with a rather regular T-shape, the confinement effects on the hard discs become negligible, leading to surface densities closer to the expected bulk value ($\eta\approx 0.54$) \cite{meyer_jamming_2010}. This is caused by a vanishing perimeter-to-surface ratio when the region's area increase, such as for a large plaza \cite{baars_random_2016}.

Further developments of this simple and versatile approach, beside the above mentioned addition of obstacles and couples, could account for pedestrian crowd motion or evacuation dynamics \cite{helbing_simulating_2000,silverberg_collective_2013,cristiani_multiscale_2014,vermuyten_review_2016}. Such computations could be a second simulation step that uses our current final configuration as a seed. We note that these extensions will most likely lack the direct and simple applicability of the approach presented here. However, we encourage experts on such techniques to promptly contact us. Collaboration can start towards, for instance, a rational design of evacuation routes for crowds constrained to social distancing. Ensuring (and accounting for) escape routes should be the next priority for development.

\begin{figure}
  \centering
  \includegraphics[width=0.5\textwidth]{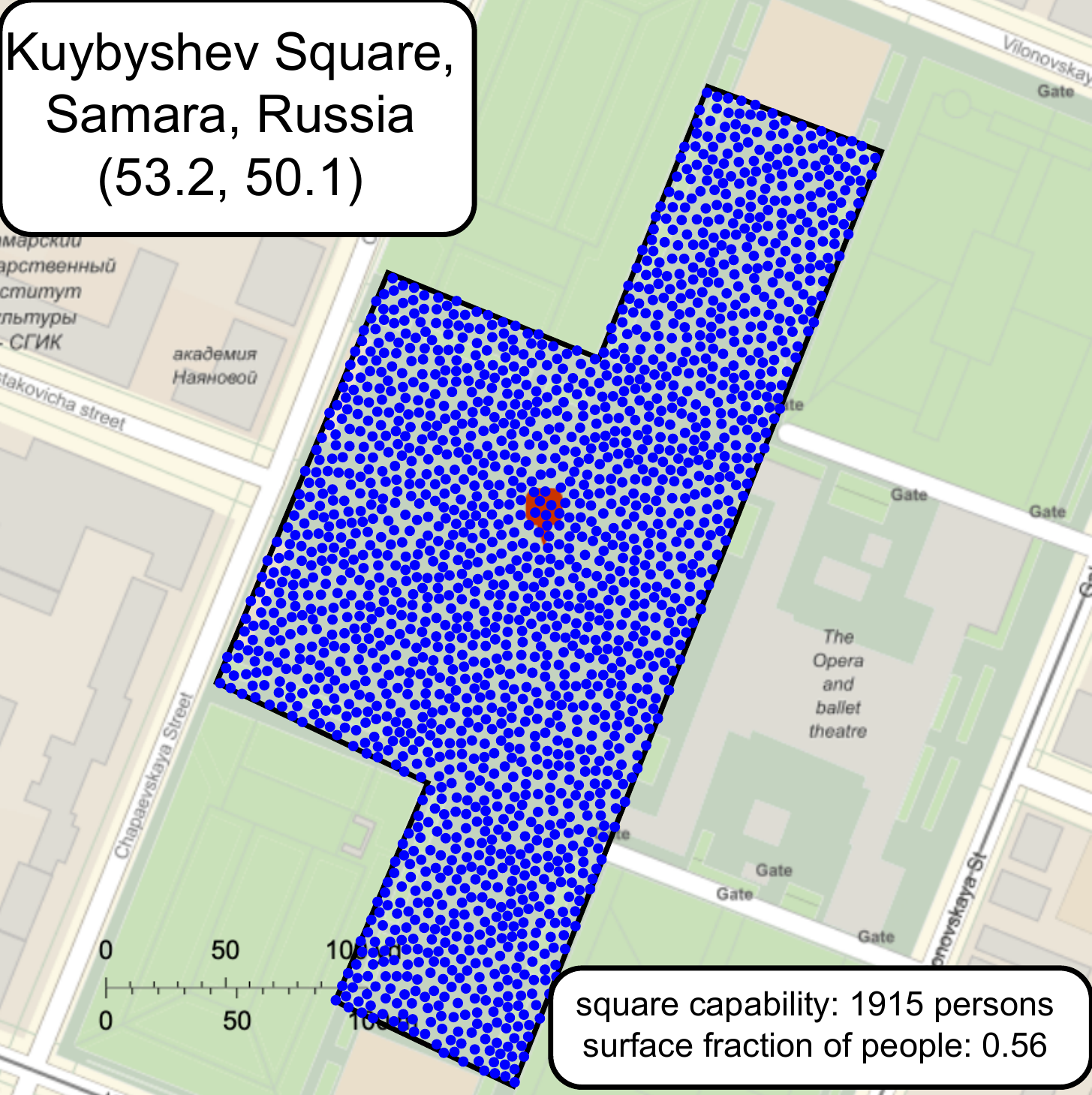}
  \caption{RSA constrained to people respecting social distance in Kuybyshev Square, a square in Samara, Russia. This is the largest square in Europe \cite{noauthor_kuibyshev_nodate}. Upon an extremely large number of person insertion attemps ($2\cdot 10^{6}$), the VLRP leads to a surface fraction of $\eta \approx 0.53$, in agreement with the expected value generated by RSA reported by Torquato\cite{torquato_random_2006}. Note $\eta$ for the VLRP is significantly lower than for the so-called random loose ($\eta\approx 0.77$) and random close ($\eta \approx 0.82$) packings \cite{meyer_jamming_2010,frost_simulation_1993}: we do not aim for packing efficiency, but for the safety of the people (freedom to move and escape from each other's trajectories).
 }
 \label{fig:fig3}
\end{figure}

In short, we provide rational guidelines for taking back our public spaces using a simple and versatile method. During the digital event HackingCovid (16-18 May 2020) \cite{ayudadig_httpshackathonayudadigitalorg_nodate}, organized by the non-profit organization AyudaDigital\cite{anahi_inicio_nodate}, we intend to develop a full open-access code of the present script and release it as an open source application as soon as possible, inspired by a similar 2D particle packing app\cite{noauthor_geyopp_nodate}. This will enable the immediate application of the proposed method. Albeit we recognize the limitation of our method up-to-date, in particular disregarding the presence of obstacles, work is ongoing in extending the procedure to take them into account. Finally, our method is not only applicable to open spaces: it could also optimize, for instance, the distribution of employees in open offices and in research labs based on the floor plans. As per today, we think its appeal, rational design, simplicity, and versatility provides very useful guidelines in the process of adapting our public (and perhaps private) spaces to the so-called 'new normality' \cite{noauthor_coronavirus_nodate}. 

\section*{Author contributions}
AGG proposed the research and built the preliminary model. AC improved the model and conducted calculations. JM conducted calculations. All authors co-wrote the manuscript. 

\section*{Acknowledgments}
We thank Cristina Hern{\'a}ndez del {\'A}lamo (Pisando Colores) and Maite Giordano (Maite de Papel) for their field work as craftswomen: they provided us with the dimensions of the vending stands and the separation distance between them. They also conducted sanity checks in our case study for Plaza de la Luz. We thank our friends and colleagues Sofia Bimpikou, Leon Bremer, Remco Tuinier, Willem K. Kegel, Martin F. Haase, Dominique Thies--Weesie, Vanessa Yanes Est{\'e}vez, and Catarina Esteves for checking the preliminary version of this manuscript and providing feedback on the feasibility and future implementations of the method. AGG thanks the AyudaDigital community for the utterly inspiring social interactions leading to this idea.

\section*{Conflict of interest}
We prefer keeping our conflicts to ourselves.

\section*{Supplemental information}
\begin{itemize}
  \item Further analysis on the effect container size and shape in the VLRP surface fraction.
 \item Auto-correlation function of the acceptance probability (from the value of the insertion (0 or 1) as a function of the number of insertion trials.
  \item Mathematica script.
\end{itemize}

\bibliographystyle{unsrt}
\bibliography{references}

\begin{thebibliography}{10}

\bibitem{noauthor_australias_2020}
Australia's coronavirus restrictions are easing — here's what you can and
  can't do - {ABC} {News}, April 2020.
\newblock Library Catalog: www.abc.net.au.

\bibitem{correspondent_coronavirus_2020}
Jon Henley~Europe correspondent.
\newblock Coronavirus 'under control' in {Germany}, as some countries plan to
  relax lockdowns.
\newblock {\em The Guardian}, April 2020.

\bibitem{welle_wwwdwcom_coronavirus_nodate}
Deutsche Welle~(www.dw.com).
\newblock Coronavirus: {Lifting} lockdowns, {European} countries go their own
  way {\textbar} {DW} {\textbar} 28.04.2020.
\newblock Library Catalog: www.dw.com.

\bibitem{who_2020}
World~Health Organization.
\newblock Advice for public.
\newblock
  \url{https://www.who.int/emergencies/diseases/novel-coronavirus-2019/advice-for-public}.
\newblock Accessed: 2020-05-10.

\bibitem{YoutubeVideo}
Youtube Video.
\newblock May {Day} {Protests} in {Lockdown}: {From} {Muted} {Celebrations} to
  {Violence}.
\newblock \url{https://www.youtube.com/watch?v=_MrKVzrW_FQ}.

\bibitem{cheng_severe_2007}
Vincent C.~C. Cheng, Susanna K.~P. Lau, Patrick C.~Y. Woo, and Kwok~Yung Yuen.
\newblock Severe {Acute} {Respiratory} {Syndrome} {Coronavirus} as an {Agent}
  of {Emerging} and {Reemerging} {Infection}.
\newblock {\em Clin. Microbiol. Rev.}, 20(4):660--694, October 2007.

\bibitem{lekkerkerker_colloids_2011}
H.~N.~W. Lekkerkerker and R.~Tuinier.
\newblock {\em Colloids and the {Depletion} {Interaction}}.
\newblock Lecture {Notes} in {Physics}. Springer The Netherlands, 2011.

\bibitem{noauthor_plotting_nodate}
plotting - {Interactively} extract points from a plot ({ListPlot} or
  {SmoothDensityHistogram}).
\newblock Library Catalog: mathematica.stackexchange.com.

\bibitem{zhang_precise_2013}
G.~Zhang and S.~Torquato.
\newblock Precise {Algorithm} to {Generate} {Random} {Sequential} {Addition} of
  {Hard} {Hyperspheres} at {Saturation}.
\newblock {\em Phys. Rev. E}, 88(5):053312, November 2013.
\newblock arXiv: 1402.4883.

\bibitem{torquato_random_2006}
S.~Torquato, O.~U. Uche, and F.~H. Stillinger.
\newblock Random sequential addition of hard spheres in high {Euclidean}
  dimensions.
\newblock {\em Phys. Rev. e}, 74(6):061308, December 2006.

\bibitem{chen_random_2017}
Elizabeth~R. Chen and Miranda Holmes-Cerfon.
\newblock Random {Sequential} {Adsorption} of {Discs} on {Surfaces} of
  {Constant} {Curvature}: {Plane}, {Sphere}, {Hyperboloid}, and {Projective}
  {Plane}.
\newblock {\em J. Nonlinear Sci.}, 27(6):1743--1787, December 2017.

\bibitem{widom_random_1966}
B.~Widom.
\newblock Random {Sequential} {Addition} of {Hard} {Spheres} to a {Volume}.
\newblock {\em J. Chem. Phys.}, 44(10):3888--3894, May 1966.

\bibitem{mathematica_2020}
Wolfram~Research{,} Inc.
\newblock Mathematica, {V}ersion 12.1.
\newblock Champaign, IL, 2020.

\bibitem{totrov_contour-buildup_1996}
Maxim Totrov and Ruben Abagyan.
\newblock The {Contour}-{Buildup} {Algorithm} to {Calculate} the {Analytical}
  {Molecular} {Surface}.
\newblock {\em J. Struct. Biol.}, 116(1):138--143, January 1996.

\bibitem{ertl_fast_2000}
Peter Ertl, Bernhard Rohde, and Paul Selzer.
\newblock Fast {Calculation} of {Molecular} {Polar} {Surface} {Area} as a {Sum}
  of {Fragment}-{Based} {Contributions} and {Its} {Application} to the
  {Prediction} of {Drug} {Transport} {Properties}.
\newblock {\em J. Med. Chem.}, 43(20):3714--3717, October 2000.

\bibitem{vorobjev_sims_1997}
Y.N. Vorobjev and J.~Hermans.
\newblock {SIMS}: computation of a smooth invariant molecular surface.
\newblock {\em Biophys. J.}, 73(2):722--732, August 1997.

\bibitem{moon_fast_1989}
Joseph~B Moon and W~Jeffrey Howe.
\newblock A fast algorithm for generating smooth molecular dot surface
  representations.
\newblock {\em J. Mol. Graphics}, 7:4, 1989.

\bibitem{gray_theory_2011}
Christopher~G. Gray, Keith~E. Gubbins, and Christopher~G. Joslin.
\newblock {\em Theory of molecular fluids. {Vol}. 2: {Applications}}.
\newblock Number~10 in International series of monographs on chemistry. Oxford
  Univ. Press, Oxford, 2011.
\newblock OCLC: 838855142.

\bibitem{garcia_polymer-mediated_2019}
{\'A}lvaro Gonz{\'a}lez~Garc{\'i}a.
\newblock {\em Polymer-{Mediated} {Phase} {Stability} of {Colloids}}.
\newblock Springer {Theses}. Springer International Publishing, 2019.

\bibitem{dorta_macarena_nodate}
Leticia Dorta.
\newblock Macarena {Fuentes}: "{Reabrimos} el mercadillo porque garantizamos la
  seguridad" {\textbar} {Daute} {Digital}.
\newblock Library Catalog: dautedigital.es.

\bibitem{noauthor_maite_nodate}
Maite de papel (@maitedepapel) • {Instagram} photos and videos.
\newblock Library Catalog: www.instagram.com.

\bibitem{noauthor_pisando_nodate}
Pisando {Colores} (@pisandocolores) • {Instagram} photos and videos.
\newblock Library Catalog: www.instagram.com.

\bibitem{meyer_jamming_2010}
Sam Meyer, Chaoming Song, Yuliang Jin, Kun Wang, and Hernán~A. Makse.
\newblock Jamming in two-dimensional packings.
\newblock {\em Physica A}, 389(22):5137--5144, November 2010.

\bibitem{baars_random_2016}
R.~J. Baars.
\newblock Random porous media and magnetic separation of magnetic colloids,
  January 2016.
\newblock Accepted: 2016-01-22T18:04:27Z ISBN: 9789039364703 Library Catalog:
  dspace.library.uu.nl Publisher: Utrecht University.

\bibitem{helbing_simulating_2000}
Dirk Helbing, Illés Farkas, and Tamás Vicsek.
\newblock Simulating dynamical features of escape panic.
\newblock {\em Nature}, 407(6803):487--490, September 2000.

\bibitem{silverberg_collective_2013}
Jesse~L. Silverberg, Matthew Bierbaum, James~P. Sethna, and Itai Cohen.
\newblock Collective {Motion} of {Humans} in {Mosh} and {Circle} {Pits} at
  {Heavy} {Metal} {Concerts}.
\newblock {\em Physical Review Letters}, 110(22):228701, May 2013.

\bibitem{cristiani_multiscale_2014}
Emiliano Cristiani, Benedetto Piccoli, and Andrea Tosin.
\newblock {\em Multiscale {Modeling} of {Pedestrian} {Dynamics}}, volume~12 of
  {\em {MS}\&{A}}.
\newblock Springer International Publishing, Cham, 2014.

\bibitem{vermuyten_review_2016}
Hendrik Vermuyten, Jeroen Beliën, Liesje De~Boeck, Genserik Reniers, and Tony
  Wauters.
\newblock A review of optimisation models for pedestrian evacuation and design
  problems.
\newblock {\em Saf. Sci.}, 87:167--178, August 2016.

\bibitem{noauthor_kuibyshev_nodate}
Kuibyshev {Square} in {Samara} - description, photo, routes, map.
\newblock Library Catalog: geomerid.com.

\bibitem{frost_simulation_1993}
R.~Frost, J.C. Schön, and P.~Salamon.
\newblock Simulation of random close packed discs and spheres.
\newblock {\em Comp. Mat. Sci.}, 1(4):343--350, October 1993.

\bibitem{ayudadig_httpshackathonayudadigitalorg_nodate}
{Ayuda Digital- A non-profit organization}.
\newblock https://hackathon.ayudadigital.org/.
\newblock Library Catalog: hackathon.ayudadigital.org.

\bibitem{anahi_inicio_nodate}
Ayuda Digital-A non-profit organization.
\newblock Ayuda digital covid - homepage.
\newblock Library Catalog: www.ayudadigitalcovid.org.

\bibitem{noauthor_geyopp_nodate}
{GeYOPP}: {Generate} your own particle packing.
\newblock Library Catalog: apps.apple.com.

\bibitem{noauthor_coronavirus_nodate}
Coronavirus part of new normality, says {German} agency as briefings cease
  {\textbar} {World} news {\textbar} {The} {Guardian}.

\end{thebibliography}


\end{document}